\documentclass[10pt]{article}

\usepackage{amsmath}
\usepackage{amssymb}
\usepackage[ruled]{algorithm2e}

\def\>{\ensuremath{\rangle}}
\def\<{\ensuremath{\langle}}

\begin{document}
\newtheorem{definition}{Definition}
\newtheorem{lemma}{Lemma}
\newtheorem{theorem}{Theorem}
\newcommand{\bra}[1]{\langle #1|}
\newcommand{\ket}[1]{|#1\rangle}
\newcommand{\braket}[3]{\langle #1|#2|#3\rangle}
\newcommand{\ip}[2]{\langle #1|#2\rangle}
\newcommand{\op}[2]{|#1\rangle \langle #2|}

\newcommand{\tr}{{\rm tr}}
\newcommand {\spa } {{\rm span}}
\newcommand {\supp } {{\rm supp}}
\newcommand{\B}{\mathcal{B}}
\newcommand {\C } {{\mathbf{C}}} %The set of complex numbers
\newcommand {\D } {{\mathcal{D}}}
\newcommand {\E } {{\mathcal{E}}}
\newcommand{\hs}{\mathcal{H}}
\newcommand {\M} {{\mathcal{M}}}
\newcommand {\T} {{\mathcal{T}}}

\title{Termination of Nondeterministic Quantum Programs}

\author{Yangjia Li, Nengkun Yu, and Mingsheng Ying\\ \\TNLIST, Dept. of CS, Tsinghua
University\\ QCIS, FEIT, University of Technology, Sydney\\liyangjia@gmail.com}
\date{}

\maketitle

\begin{abstract}
We define a language-independent model of nondeterministic quantum
programs in which a quantum program consists of a finite set of
quantum processes. These processes are represented by quantum Markov
chains over the common state space. An execution of a
nondeterministic quantum program is modeled by a sequence of actions
of individual processes. These actions are described by
super-operators on the state Hilbert space. At each step of an
execution, a process is chosen nondeterministically to perform the
next action.

A characterization of reachable space and a characterization of
diverging states of a nondeterministic quantum program are
presented. We establish a zero-one law for termination probability
of the states in the reachable space of a nondeterministic quantum
program. A combination of these results leads to a necessary and
sufficient condition for termination of nondeterministic quantum
programs. Based on this condition, an algorithm is found for
checking termination of nondeterministic quantum programs within a
fixed finite-dimensional state space.

A striking difference between nondeterministic classical and quantum programs is shown by example: it is possible that each of several quantum programs simulates the same classical program which terminates with probability 1, but the nondeterministic program consisting of them terminates with probability 0 due to the interference carried in the execution of them.
\end{abstract}

\section{Introduction}

Quantum algorithms are usually expressed at the very low-level of
quantum circuits. As pointed out by Abramsky~\cite{Ab04},
high-level, conceptual methods are needed for designing, programming
and reasoning about quantum computational systems. Along this line,
intensive research on quantum programming has been conducted in the
last 15 years. Several quantum programming languages have been
defined, including QCL by \"{O}mer~\cite{Om03}, qGCL by Sanders and
Zuliani~\cite{SZ00}, a quantum extension of C++ by Betteli et
al.~\cite{BCS03}, QPL by Selinger~\cite{Se04}, and QML by Altenkirch
and Grattage~\cite{AG05}. The operational or denotational semantics
of these languages have been introduced. D'Hondt and
Panangaden~\cite{DP06} proposed the notion of quantum weakest
precondition, and then a predicate transformer semantics of quantum
programs was presented in~\cite{YDFJ10}. Also, several proof systems
for verification of quantum programs have been developed~\cite{BS06,
BJ04, CMS06, FDJY07, Yi11}, and some approaches to the
implementation of quantum programming languages have been
suggested~\cite{NPW07, SA06, Zu05, YF11}. Furthermore, several
quantum process algebras have been proposed: CQP by Gay and
Nagarajan~\cite{GN05}, QAlg by Jorrand and Lalire~\cite{JL04} and
qCCS~\cite{FDY11}, to model quantum communication and concurrency.
For a more systematic exposition, we refer to two excellent survey
papers~\cite{G06, Se04a}.

Nondeterminism provides an important high-level feature in classical
computation for specifying programs' behavior, without having to
specify details of implementations. Zuliani~\cite{Zu04} found a way
for embedding nondeterminism into his quantum programming language
qGCL, and then used qGCL equipped with a nondeterministic choice
constrct to model and reason about Mitchison and Josza's
counterfactual computation~\cite{MJ01} and quantum systems in mixed
states. In this paper, we consider a class of nondeterministic
quantum programs defined in a language-independent way. A
nondeterministic quantum program consists of a collection of quantum
processes. These processes are described by quantum Markov chains
over the common state space. This model of nondeterministic quantum
programs is indeed a quantum generalization of Markov decision
processes, which are widely used in the studies of probabilistic
programs, see for example~\cite{HSP83}.

This paper focuses on the termination problem of nondeterministic
quantum programs within a fixed finite-dimensional state space. The
paper is organized as follows. In Sec.~\ref{Pre}, we briefly review
the basic notions from quantum theory required in this paper, with
an emphasis on fixing notations. In Sec.~\ref{Mod}, a model of
nondeterministic quantum programs is defined in terms of quantum
Markov chains. In this model, an execution of a nondeterministic
quantum program is a sequence of actions of individual processes,
and following Selinger~\cite{Se04} these actions are depicted by
super-operators on the state Hilbert space. At each step of an
execution, a process is chosen nondeterministically to perform the
next action. We define the termination probability of a
nondeterministic quantum program starting in a state according to an
execution schedule. Then the termination of a nondeterministic
quantum program is defined to be the infimum of its termination
probabilities over all possible schedules. At the end of this
section, we consider an example of nondeterministic quantum program
consisting of two quantum walks on a graph~\cite{AAKV01}. This
example is interesting because it indicates a striking difference between nondeterministic classical and quantum programs: it is possible that each of several quantum programs simulates the same classical program which terminates with probability 1, but the nondeterministic program consisting of them terminates with probability 0 due to the interference carried in the execution of them. In Sec.~\ref{RS}, we examine the reachable space of a nondeterministic quantum program. By taking the
arithmetic average of the super-operators performed by individual
processes, we are able to define a deterministic quantum program
whose reachable space is equal to the reachable space of the
original nondeterministic program. Furthermore, the reachable space
of the average deterministic program can be obtained by recursively
constructing a finite increasing sequence of subspaces of the state
Hilbert space. The notions of terminating and diverging states of a
nondeterministic quantum program are introduced in Sec.~\ref{TS}.
The structures of the sets of terminating and diverging states are
clarified. In particular, it is shown that the space of diverging
pure states can also be recursively constructed in a finite number
of steps. In Sec.~\ref{ZO}, the Hart-Sharir-Pnueli zero-one law for
probabilistic concurrent programs~\cite{HSP83} is generalized to the
case of nondeterministic quantum programs. This quantum zero-one law
enables us to discover an algorithmically checkable termination
condition for nondeterministic quantum programs in terms of
reachable space and diverging pure states. A classical (not quantum)
algorithm for termination checking of nondeterministic quantum
programs is then presented in Sec.~\ref{Alg}. A brief conclusion is drawn in Sec.~\ref{Con}.

\section{Preliminaries and Notations}\label{Pre}
We assume that the reader is familiar with basic quantum theory, and
the main aim of this section is to fix notations.
\subsection{Quantum states}
In quantum mechanics, the state space of a physical system is
described by a complex Hilbert space $\hs$. In this paper, we only
consider finite-dimensional Hilbert spaces. We write $\dim\hs$ for
the dimension of space $\hs$. A pure state of a system is
represented by a unit vector in the state space of the system, and a
mixed state by a (partial) density operator, that is, a positive
semi-definite matrix $\rho$ with its trace $\tr(\rho)\leq 1$. We
write $\mathcal{D}(\hs)$ for the set of (partial) density operators
on $\hs$. For convenience, we simply write $\psi$ for the density
operator corresponding to pure state $\ket{\psi}$, that is,
$\psi=\op{\psi}{\psi}$. For any two density operators $\rho$ and
$\sigma$, their distance is defined to be
$\parallel\rho-\sigma\parallel_*$, where $$\parallel
M\parallel_*=\tr(\sqrt{MM^\dagger})$$ is the trace norm of $M$ for
all operators $M$. Let $$\rho=\sum_{i}\lambda_i\op{\psi_i}{\psi_i}$$
be the spectral decomposition of density operator $\rho$. The
subspace $\spa\{\ket{\psi_i}\}$ is called the support of $\rho$,
written $\supp \rho$. Recall that for a family $\{X_i\}$ of
subspaces of $\hs$, the join of $\{X_i\}$ is defined by $$\bigvee_i
X_i=\spa(\bigcup_i X_i),$$ and we write $X\vee Y$ for the join of
two subspaces $X$ and $Y$. Then it is easy to verify that for any
two states $\rho,\sigma\in\D(\hs)$,
$$\supp(\rho+\sigma)=\supp(\rho)\vee\supp(\sigma).$$

\subsection{Quantum operations}
Super-operators formalize physical transformations between quantum
states. A super-operator $\E$ on a Hilbert space $\hs$ is a linear
map from linear operators on $\hs$ to themselves satisfying the
following two conditions: \begin{enumerate}\item Completely
positive: for any a extra state space $\hs'$ and any positive
semi-definite operator $P\in\D(\hs'\otimes\hs)$,
$(\mathcal{I}\otimes\E)(P)$ is always positive semi-definite, where
$\mathcal{I}:\B(\hs')\rightarrow\B(\hs')$ is the identity
super-operator; \item Trace-preserving: for any $\rho\in\D(\hs)$,
$\tr(\E(\rho))=\tr(\rho)$.\end{enumerate} The Kraus representation
theorem asserts that a linear map $\E$ is a super-operator iff there
are linear operators $E_i$ such that $$\E(\rho)=\sum_{i} E_i\rho
E_i^\dag$$ for all $\rho\in\mathcal{D}(\hs)$, and $$\sum_i E_i^\dag
E_i=Id_\hs,$$ where $Id_\hs$ is the identity operator on $\hs$.

For any subspace $X$ of $\hs$, we define the image of $X$ under $\E$
as $$\E(X)=\bigvee_{\ket{\psi}\in X}\supp \E(\psi).$$ In other
words, if we write $P_X$ for the projection operator of $X$, then
$\E(X)=\supp\E(P_X)$. Using the Kraus representation, it is easy to
verify that $\E(\supp(\rho))=\supp \E(\rho)$. We can also define the
pre-image of $X$ under $\E$ by $$\E^{-1}(X)=\{\ket{\psi}\in\hs|\supp
\E(\psi)\subseteq X\}.$$ It is actually the maximal subspace $Y$
satisfying that $\E(Y)\subseteq X$. We write $X^\perp$ for the orthogonal complement of a subspace $X$, then it is also easy to verify that
$$\E^{-1}(X)=(\E^*(X^\perp))^\perp,$$
where the super-operator
$$\E^*(\cdot)=\sum_i E_i^\dagger\cdot E_i$$
is the Schr\"{o}dinger-Heisenberg dual of
$$\E(\cdot)=\sum_i E_i\cdot E_i^\dagger.$$

\subsection{Quantum measurements}
To acquire information about a quantum system, a measurement must be
performed on it. A quantum measurement on a system with state space
$\hs$ is described by a collection $\{M_i\}$ of linear operators on
$\hs$ satisfying $$\sum_i M_i^\dag M_i=Id_\hs,$$ where indices $i$
stand for the outcomes that may occur in the experiment. If the
system is in state $\rho\in\mathcal{D}(\hs)$ immediately before the
measurement, then the probability that result $i$ occurs is
$\tr(M_i\rho M_i^\dagger)$, and the state of the system after the
measurement is $M_i\rho M_i^\dagger$.

\section{A Model of Nondeterministic Quantum Programs}\label{Mod}
\subsection{Basic Definitions}
\begin{definition}Let $\hs$ be a finite-dimensional Hilbert space
which will be used as the state space of programs. A
nondeterministic quantum program is a pair
$$\mathcal{P}=(\{\E_i:i=1, \cdots, m\}, \{M_0,M_1\}),$$ where:
\begin{enumerate}
\item $\E_i$ is a super-operator on $\hs$ for each $i=1,\cdots,m$;
\item $\{M_0, M_1\}$ is a measurement on $\hs$.
\end{enumerate}
\end{definition}

There are $m$ processes in the program $\mathcal{P}$. The one-step
running of process $i$ is modeled by super-operator $\E_i$ for each
$1\leq i\leq m$. We will call $\mathcal{P}$ a deterministic quantum
program when $m=1$, that is, there is only one process in
$\mathcal{P}$.

We now see how a nondeterministic quantum program be executed. We
first consider a single computation step of program $\mathcal{P}$.
It is achieved as follows:
\begin{itemize}
\item Before each step, the measurement $\{M_0,M_1\}$ is performed on the
current
 state $\rho$ to determine whether the program terminates or not. If the outcome is $0$, the program terminates;
 Otherwise the program goes to complete a step then.
\item In each step, an element $i$ is nondeterministically chosen from the index set $\{1,2,\cdots ,m\}$
firstly, and then the operation $\E_i$ is performed on the current
program state. Thus, the state becomes $\E_i(M_1\rho M_1^\dagger)$
after the measurement and the operation $\E_i$.
\end{itemize}

A computation of a nondeterministic quantum program is a finite or
infinite sequence of computation steps in which the same measurement
is performed to determine termination of the program in all steps,
but the super-operators performed in different steps are usually
different and they are nondeterministically scheduled. Formally, the
set of schedules of program $\mathcal{P}$ is defined to be
\begin{equation*}\begin{split}S&=\{1,2,\cdots,m\}^\infty\\ &=\{s_1 s_2\cdots
s_k\cdots:s_k\in\{1,2,\cdots,m\}\ {\rm for\ all}\ k\geq
0\}.\end{split}\end{equation*} We also define the set of schedule
fragments of $\mathcal{P}$ to be
$$S_{fin}=\{1,2,\cdots,m\}^*=\bigcup_{n=0}^{\infty}\{1,2,\cdots,m\}^n.$$
For convenience, we use $\epsilon$ to represent empty string. For
any $f=s_1\cdots s_m\in S_{fin}$, we write $|f|$ for the length of
$f$, that is, $|f|=m$. For each $n\leq |f|$, $f(\leq n)$ stands for
the head $s_1\cdots s_n$ of $f$. We also write the head $$s(\leq
n)=s_1s_2\cdots s_n\in S_{fin}$$ and the tail
$$s(>n)=s_{n+1}s_{n+2}\cdots\in S$$ of $s=s_1s_2\cdots\in S$ for each
$n\geq 0$. For any $s=s_1s_2\cdots\in S$ and $f=s_1^\prime\cdots
s_m^\prime\in S_{fin}$, we write $fs$ for the concatenation of $f$
and $s$, that is, the schedule $$fs=s_1^\prime\cdots s_m^\prime
s_1s_2\cdots.$$ For simplicity of presentation, we introduce the
notation $\mathcal{T}_i$ which stands for the super-operator defined
by $$\mathcal{T}_{i}(\rho)=\E_i(M_1\rho M_1^\dagger)$$ for every
$\rho\in\D (\mathcal{H})$ and $1\leq i\leq m$. Furthermore, for
any $f=s_1s_2\cdots s_n\in S_{fin}$, we write:
$$\mathcal{T}_f=\mathcal{T}_{s_n}\circ\cdots\circ
\mathcal{T}_{s_2}\circ \mathcal{T}_{s_1},$$ in particular,
$\mathcal{T}_\epsilon(\rho)=\rho$ for all $\rho$. Let
$\rho\in\mathcal{D}(\mathcal{H})$. If the input is state $\rho$, and
program $\mathcal{P}$ is executed according to a schedule
$s=s_1s_2\cdots \in S$, the program state after $n$ steps is
$\mathcal{T}_{s(\leq n)}(\rho)$.

\subsection{Termination Probability}

Suppose that the input state to program $\mathcal{P}$ is $\rho$. For
any schedule fragment $f\in S_{fin}$, we define the probability that
the program terminates within $f$ as follows:
$$t_f(\rho)=\sum\limits_{n=0}^{|f|}\tr(M_0\mathcal{T}_{f(\leq n)}(\rho)M_0^\dagger).$$
If the program is executed according to a schedule $s=s_1s_2\cdots$,
it is easy to see that the probability of the program terminating in
no more than $n$ steps is $t_{s(\leq n)}(\rho)$. Furthermore, the
probability that the program terminates in a finite number of steps
is
$$t_s(\rho)=\lim_{n\to\infty}t_{s(\leq n)}(\rho)=\sum\limits_{n=0}^\infty\tr(M_0\mathcal{T}
_{s(\leq n)}(\rho)M_0^\dagger).$$ It is obvious that $tr(\rho)\geq
t_s(\rho)$, and $tr(\rho)-t_s(\rho)$ is the divergence probability
of the program starting in state $\rho$ and executed according to
schedule $s$. We can divide the termination probability $t_s(\rho)$
into two parts:
\begin{itemize}\item The first part is the probability of
terminating in less than $n$ steps, that is
\begin{equation}\label{equ:first-n}\begin{split}
t_{s(\leq n-1)}(\rho)&=\sum_{k=0}^{n-1}\tr(M_0\mathcal{T}_{s(\leq
k)}(\rho)M_0^\dagger)\\ &=\tr(\rho)-\tr(M_1\mathcal{T}_{s(\leq
n-1)}(\rho)M_1^\dagger)\\ &=\tr(\rho)-\tr(\mathcal{T}_{s(\leq
n)}(\rho))
\end{split}\end{equation}
because all $\mathcal{E}_i$ $(i=1,...,m)$ are trace-preserving.
\item The second part is the probability of terminating
in at least $n$ steps, that is
\begin{equation}\label{equ:n-left}
\begin{split}&\sum_{k=n}^{\infty}\tr(M_0\mathcal{T}_{s(\leq
k)}(\rho)M_0^\dagger)\\
&=\sum_{k=0}^{\infty}\tr(M_1(\mathcal{T}_{s_{n+1}s_{n+2}\cdots
s_{n+k}} \circ \mathcal{T}_{s(\leq n)})(\rho)M_1^\dagger)\\
&=t_{s(>n)}(\mathcal{T}_{s(\leq n)}(\rho)).\end{split}
\end{equation}\end{itemize}
Combining the above two equations, we get that
\begin{equation}\label{equ:two-part}\tr(\rho)-t_s(\rho)=\tr(\mathcal{T}_{s(\leq n)}(\rho))-t_{s(>n)}(\mathcal{T}_{s(\leq
n)}(\rho)).
\end{equation}This indicates that the divergence probability of a program is an
invariant through an execution path of the program.

In general, an execution along with any schedule $s\in S$ is
possible for a nondeterministic program. So, we need to consider all
possible execution paths of the program together.

\begin{definition}The termination probability of program
$\mathcal{P}$ starting in state $\rho$ is
$$t(\rho)=\inf\{t_s(\rho)|s\in S\}.$$
\end{definition}

\subsection{An Example: Quantum Walks}\label{sub:qw}

We consider quantum walks on a graph. Let $C_4=(V,E)$ be a circle
with four vertices, where $V=\{0,1,2,3\}$ is the set of vertices,
and $E=\{(0,1), (1,2),(2,3),(3,0)\}$ is the set of edges. We first
define a quantum walk on $C_4$ as follows:
\begin{itemize}
\item The state Hilbert space is $\C^V$, and it has $\{\ket{i}|i\in
V\}$ as its computational basis;
\item The initial state is $\ket{0}$. This means that the walk start at the vertex $0$;
\item A single step of the walk is defined by the unitary operator:
$$W_1=\frac{1}{\sqrt{3}}\left(\begin{array}{cccc}
1 & 1 & 0 & -1\\
1 & -1 & 1 & 0\\
0 & 1 & 1 & 1\\
1 & 0 & -1 & 1
\end{array}\right).$$
It means that at any vertex, the probability of walking to the left
and walking to the right are both $1/3$, and there is also a
probability $1/3$ of not walking.
\item The termination measurement $\{P_0,P_1\}$ is defined by $$P_0=\op{2}{2},\ P_1=Id_4-\op{2}{2},$$ which means that there is an absorbing boundary at vertex $2$. Here,
 $Id_4$ is the $4\times 4$ unit matrix.
\end{itemize}
This quantum walk can be seen as a deterministic quantum program
$(W_1,\{P_0,P_1\})$ starting in state $\ket{0}$. It is easy to verify
that this program terminates with probability $1$, that is,
$t(\op{0}{0})=1$. If the unitary operator $W_1$ in the above quantum
walk is replaced by:
$$W_2=\frac{1}{\sqrt{3}}\left(\begin{array}{cccc}
1 & 1 & 0 & 1\\
-1 & 1 & -1 & 0\\
0 & 1 & 1 & -1\\
1 & 0 & -1 & -1
\end{array}\right),$$
then we get a new quantum walk, which can also be seen as a
deterministic quantum program $(W_2,\{P_0,P_1\})$ starting in state
$\ket{0}$. This new quantum walk is terminating too. However, if we
combine these two walks to form a nondeterministic quantum program
$(\{W_1,W_2\},\{P_0,P_1\})$, then it is not terminating when
starting in state $\ket{0}$. In fact, since $W_2W_1\ket{0}=\ket{0}$,
it holds that $$t(\ket{0}\langle 0|)\leq t_s(\ket{0}\langle 0|)=0$$
for the infinite execution path $s=(12)^\infty=121212\cdots$.

\section{Reachable Space}\label{RS}

From now on, we consider a fixed nondeterministic quantum program
$$\mathcal{P}=(\{\E_i:i=1, \cdots, m\}, \{M_0,M_1\}).$$

\begin{definition}\begin{enumerate}\item The set of reachable states of program $\mathcal{P}$
starting in state $\rho$ is $$R(\rho)=\{\mathcal{T}_f(\rho)|f\in
S_{fin}\}.$$\item The reachable space of program $\mathcal{P}$
starting in state $\rho$ is the subspace of $\hs$ spanned by
$R(\rho)$, that is,
$$\hs_{R(\rho)}=\bigvee\{\supp\sigma|\sigma\in R(\rho)\}.$$
\end{enumerate}\end{definition}

We imagine that during the running of $\mathcal{P}$, if each
nondeterministic choice of $i\in\{1,2,\cdots,m\}$ is made according
to the uniform probability distribution, then then $\mathcal{P}$
actually implements a deterministic quantum program, which can be
described as:
\begin{definition}
The average of $\mathcal{P}$ is the deterministic quantum program
$$\overline{\mathcal{P}}=(\E,\{M_0,M_1\}),$$ where $\{M_0,M_1\}$ is
the same as in $\mathcal{P}$, and $\E$ is the arithmetic average of
$\E_1,\E_2,\cdots,\E_m$, that is,
$$\E(\rho)=\frac{1}{m}\sum_{i=1}^m\E_i(\rho)$$ for every
$\rho\in\mathcal{D}(\mathcal{H})$.
\end{definition}

We define: $$\mathcal{T}(\rho)=\E(M_1\rho M_1^\dag)$$ for every
$\rho\in\mathcal{D}(\mathcal{H})$. Then the reachable set and
reachable space of $\overline{\mathcal{P}}$ starting in state $\rho$
are, respectively:\begin{equation*}\begin{split}
\overline{R}(\rho)&=\{\mathcal{T}^n(\rho)|n=0,1,2,...\},\\
\mathcal{H}_{\overline{R}(\rho)}&=\bigvee\{\supp
(\sigma)|\sigma\in\overline{R}(\rho)\}.
\end{split}\end{equation*}
The following lemma shows that $\mathcal{P}$ and its average
$\overline{\mathcal{P}}$ have the same reachable space.
\begin{lemma}\label{rea} For any $\rho\in\mathcal{D}(\mathcal{H})$,
$\hs_{R(\rho)}=\hs_{\overline{R}(\rho)}$.
\end{lemma}
\textit{Proof:} For each $n\geq 0$, we have:
\begin{eqnarray*}\mathcal{T}^n(\rho) &=&(\frac{1}{m}\sum_{i=1}^m \mathcal{T}_i)^n(\rho)\\
&=&\frac{1}{m^n}\sum_{s_1,s_2,\cdots,s_n\in S}
(\mathcal{T}_{s_n}\circ\cdots\circ \mathcal{T}_{s_2}\circ
\mathcal{T}_{s_1})(\rho)\\
&=&\frac{1}{m^n}\sum_{f\in
S_{fin},|f|=n}\mathcal{T}_f(\rho).\end{eqnarray*} Then the proof is
completed by observing that
\begin{equation*}\begin{split}\mathcal{H}_{R(\rho)}&=
\bigvee_{f\in S_{fin}}\supp(\mathcal{T}_f(\rho))\\
&=\bigvee_{n=0}^\infty\bigvee_{f\in S_{fin},|f|=n}
\supp(\mathcal{T}_f(\rho))\\ &=\bigvee_{n=0}^\infty\supp
(\mathcal{T}^n(\rho))=\mathcal{H}_{\overline{R}(\rho)}.\
\Box\end{split}\end{equation*}

Now we only need to examine the reachable space
$\mathcal{H}_{\overline{R}(\rho)}$. For every $n\geq 0$, we define
$\mathcal{H}_{\overline{R}_n(\rho)}$ to be the reachable space of
program $\overline{\mathcal{P}}$ within $n$ steps when starting in
state $\rho$, that is,
$$\mathcal{H}_{\overline{R}_n(\rho)}=\bigvee_{k=0}^n\supp(\mathcal{T}^k(\rho)).$$
Then it is clear that
$$\hs_{\overline{R}(\rho)}=\bigvee_{n=0}^\infty
\hs_{\overline{R}_n(\rho)}.$$ On the other hand, all elements of the
increasing chain
$$\hs_{\overline{R}_1(\rho)}\subseteq\hs_{\overline{R}_2(\rho)}\subseteq\cdots$$
are subspace of finite-dimensional space $\hs$. There must be some
$n\geq 0$ such that
$\hs_{\overline{R}_n(\rho)}=\hs_{\overline{R}_k(\rho)}$ for all
$k\geq n$, and thus
$\hs_{\overline{R}(\rho)}=\hs_{\overline{R}_n(\rho)}$. Furthermore,
we have a recursive characterization of reachable spaces
$\mathcal{H}_{\overline{R}_n(\rho)}$.

\begin{lemma}\label{reac}For all $n\geq 0$, we have: $$\hs_{\overline{R}_{n+1}(\rho)}=\hs_{\overline{R}_n(\rho)}
\vee \mathcal{T}(\hs_{\overline{R}_n(\rho)}).$$
\end{lemma}
\textit{Proof:} It holds that \begin{equation*}\begin{split}
\supp\mathcal{T}^{n+1}&(\rho)=\supp(\mathcal{T}(\mathcal{T}^n(\rho_n)))\\
& =\mathcal{T}(\supp(\mathcal{T}^n(\rho)))\subseteq
\mathcal{T}(\hs_{\mathcal{R}_n(\rho)}).\end{split}\end{equation*}
So, we have:
$$\hs_{\overline{R}_{n+1}(\rho)}\subseteq\hs_{\mathcal{R}_n(\rho)}\vee
\mathcal{T}(\hs_{\mathcal{R}_n(\rho)}).$$ Conversely, it holds that
\begin{equation*}\begin{split}
&\mathcal{T}(\hs_{\mathcal{R}_n(\rho)})=\mathcal{T}(\supp(\sum_{k=0}^n
\mathcal{T}^k(\rho)))\\ &= \supp(\mathcal{T}(\sum_{k=0}^n
\mathcal{T}^k(\rho)))
=\supp(\sum_{k=0}^n \mathcal{T}^{k+1}(\rho))\\
&=\bigvee_{k=1}^{n+1}\supp(\mathcal{T}^k(\rho))\subseteq\hs_{\mathcal{R}_{n+1}(\rho)}.\end{split}\end{equation*}
Thus, we have: $$\hs_{\overline{R}_n(\rho)}\vee
\mathcal{T}(\hs_{\overline{R}_n}(\rho))\subseteq\hs_{\overline{R}_{n+1}(\rho)}.\
\Box$$

Now, we are able to prove the main result in this section.
\begin{theorem}\label{thm-R} If $n$ is the smallest integer $n$ satisfying
$\hs_{\overline{R}_n(\rho)}=\hs_{\overline{R}_{n+1}(\rho)}$, then
$\hs_{R(\rho)}=\hs_{\overline{R}_n(\rho)}$.
\end{theorem}
\textit{Proof:} By Lemma~\ref{rea}, it suffices to show that
$$\hs_{\overline{R}_n(\rho)}=\hs_{\overline{R}_{n+1}(\rho)}\ {\rm
implies}\
\hs_{\overline{R}_{n+1}(\rho)}=\hs_{\overline{R}_{n+2}(\rho)},$$
which in turn implies
$\hs_{\overline{R}_{n+k}(\rho)}=\hs_{\overline{R}_{n+k+1}(\rho)}$
for all $k\geq 2$. In fact, it follows from Lemma~\ref{reac} that
\begin{equation*}\begin{split}\supp(\mathcal{T}^{n+2}&(\rho))=\supp(\mathcal{T}(\mathcal{T}^{n+1}(\rho)))\\
&=\mathcal{T} (\supp(\mathcal{T}^{n+1}(\rho)))\subseteq
\mathcal{T}(\hs_{\overline{R}_{n+1}(\rho)})\\ &
=\mathcal{T}(\hs_{\overline{R}_n}(\rho)) \subseteq
\hs_{\overline{R}_{n+1}(\rho)}.\end{split}\end{equation*} Thus, we
have
$\hs_{\overline{R}_{n+2}(\rho)}=\hs_{\overline{R}_{n+1}(\rho)}$.
$\Box$

\section{Terminating States and Diverging States}\label{TS}

\begin{definition}\begin{enumerate}\item For any $\rho\in\mathcal{D}(\mathcal{H})$, if $t(\rho)=\tr(\rho)$,
then we say that $\rho$ is a terminating state of
$\mathcal{P}$.\item We write $T$ for the set of terminating states
of $\mathcal{P}$, that is, $$T=\{\rho\in\mathcal{D}(\mathcal{H})|
t(\rho)=\tr(\rho)\}.$$\end{enumerate}
\end{definition}

The equality $t(\rho)=\tr(\rho)$ is usually called the terminating
condition of program $\mathcal{P}$. The intuitive meaning of this
condition is that whenever the program starts in state $\rho$, it
will terminate in a finite number of steps with probability $1$.
Some basic properties of terminating states are collected in the
following:

\begin{lemma}\label{lem1}\begin{enumerate}\item $\rho\in T$ iff for all $R(\rho)\subseteq T$, that is, for all $f\in
S_{fin}$, $\mathcal{T}_f(\rho)$ is a terminating state.
\item Suppose that $\rho_1,\rho_2\in\mathcal{D}(\mathcal{H})$
with $\rho_1+\rho_2\in\mathcal{D}(\mathcal{H})$. Then
$\rho_1+\rho_2\in T$ iff $\rho_1\in T$ and $\rho_2\in T$. \item Let
$\ket{\psi}$ and $\ket{\varphi}$ be two pure states. If $\ket{\psi},
\ket{\varphi}\in T$, then any pure state
$\ket{\xi}=a\ket{\psi}+b\ket{\varphi}\in T$, where $a,b\in
\mathbf{C}$.
\end{enumerate}
\end{lemma}

\textit{Proof:} \begin{enumerate} \item The ``if" part is obvious by
putting $f=\epsilon$. To prove the ``only if" part, we assume that
$\rho\in T$. Then for any $s\in S$, it follows from Eq.~(\ref{equ:two-part}) that
$$tr(\mathcal{T}_f(\rho))-t_s(\mathcal{T}_f(\rho))=tr(\rho)-\mathcal{T}_{fs}(\rho)=0.$$
The arbitrariness of $s$ implies that
$tr(\mathcal{T}_f(\rho))=t(\mathcal{T}_f(\rho)).$

\item If $\rho_1\in T$ and $\rho_2\in T$, then $t(\rho_i)=tr(\rho_i)$
$(i=1,2)$, and \begin{equation*}\begin{split}tr(\rho_1&+\rho_2)\geq
t(\rho_1+\rho_2)=\inf\{t_s(\rho_1+\rho_2)|s\in S\}\\ &
=\inf\{t_s(\rho_1)+t_s(\rho_2)|s\in S\}\geq t(\rho_1)+t(\rho_2)\\
&=tr(\rho_1)+tr(\rho_2)=tr(\rho_1+\rho_2).\end{split}\end{equation*}
So, $t(\rho_1+\rho_2)=tr(\rho_1+\rho_2)$, and $\rho_1+\rho_2\in T$.

Conversely, if $\rho_1+\rho_2\in T$, then for each $s\in S$,
\begin{equation*}\begin{split}t_s(\rho_1)+t_s(\rho_2)&=t_s(\rho_1+\rho_2)
\\ &=tr(\rho_1+\rho_2)=tr(\rho_1)+tr(\rho_2).\end{split}\end{equation*} Since $t_s(\rho_i)\leq
tr(\rho_i)$ $(i=1,2)$, it must be that $t_s(\rho_i)=tr(\rho_i)$
$(i=1,2)$. Therefore, $t(\rho_i)=tr(\rho_i)$, and $\rho_i\in T$ $(i=1,2)$.

\item Put $\ket{\xi}=a\ket{\psi}+b\ket{\varphi}$ and
$\ket{\xi'}=a\ket{\psi}-b\ket{\varphi}$. Then we have:
\begin{equation*}\begin{split}&t(\xi+\xi')=t(\ket{\xi}\langle\xi|+\ket{\xi^\prime}\langle\xi^\prime|)\\ &=
t(2|a|^2\psi+2|b|^2\varphi) =2|a|^2\tr(\psi)+2|b|^2\tr(\varphi)\\
&=\tr(\xi)+\tr(\xi')=\tr(\xi+\xi').\end{split}\end{equation*} Note
that in the above equation, we slightly abuse the notation of
density operator allowing unnormalization with trace greater than
$1$. This is not problematic because of linearity. So,
$\xi+\xi^\prime\in T$, and it follows from item 2 that $\xi\in T$.
$\Box$\end{enumerate}

\begin{definition}\begin{enumerate}\item For any $\rho\in\mathcal{D}(\mathcal{H})$,
if for some schedule $s\in S$, we have $t_s(\rho)=0$, then we say
that $\rho$ is a diverging state of $\mathcal{P}$.\item We write $D$
for the set of diverging states of $\mathcal{P}$, that is,
$$D=\{\rho\in\mathcal{D}(\mathcal{H})| t_s(\rho)=0\ {\rm for\ some}\
s\in S\}.$$ \item We write $PD$ for the diverging pure state of
$\mathcal{P}$, that is, $$PD=\{\ket{\psi}\in\hs|t_s(\psi)=0\ {\rm
for\ some}\ s\in S\}.$$\end{enumerate}
\end{definition}

The remainder of this section is devoted to examine the structure of
diverging pure states $PD$, which is crucial in developing an
algorithm for checking termination of program $\mathcal{P}$ in
Sec.~\ref{Alg}. To this end, we introduce some auxiliary notions:

\begin{definition}\begin{enumerate}\item For each schedule fragment $f\in
S_{fin}$, we define:
$$PD_f=\{\ket{\psi}\in\mathcal{H}|t_f(\psi)=0\}.$$
\item For each $n\geq 0$, we define: $$PD_n=\bigcup_{f\in S_{fin},|f|=n}PD_f.$$
\item For each schedule $s\in S$, we define:
$$PD_s=\{\ket{\psi}\in\mathcal{H}|t_s(\psi)=0\}.$$
\end{enumerate}\end{definition}

By definition, we have: $$PD=\bigcup_{s\in S}PD_s.$$ For any $s\in
S$ and $n_1\geq n_2$, It holds that $PD_{s(\leq n_1)}\subseteq
PD_{s(\leq n_2)}$ because $t_{s(\leq n_1)}\geq t_{s(\leq n_2)}$.
Furthermore, we have:
\begin{equation}\label{inter}PD_s=\bigcap_{n=0}^\infty PD_{s(\leq n)}\end{equation} since
$t_s(\cdot)=\lim_{n\to\infty}t_{s(\leq n)}(\cdot)$.

\begin{lemma}\label{lem:subspace}For any $f\in S_{fin}$, we
have:\begin{enumerate}\item $PD_f$ is a subspace of $\hs$.\item Let
$$\hs_0=\{\ket{\psi}|M_0\ket{\psi}=0\}$$ be the orthogonal
complementary subspace of measurement operator $M_0$. Then
$PD_f\subseteq\hs_0$.\item For each $k\in\{1,2,\cdots,m\}$,\begin{equation}\label{equ:recur}PD_{kf}=H_0\cap \mathcal{T}_{k}^{-1}(PD_f)\end{equation}
\end{enumerate}
\end{lemma}

\textit{Proof:}\begin{enumerate}\item Let $\ket{\psi}$,
$\ket{\varphi}$ be any two states in $PD_f$, and
$\ket{\xi}=a\ket{\psi}+b\ket{\varphi}$ be any linear superposition
of them. Write $\ket{\xi'}=a\ket{\psi}-b\ket{\varphi}$. Then
\begin{equation*}\begin{split}&t_f(\xi)+t_f(\xi')=t_f(\xi+\xi')
=t_f(2|a|^2\psi+2|b|^2\varphi)\\
&=2|a|^2t_f(\psi)+2|b|^2t_f(\varphi)=0.\end{split}\end{equation*}
Thus $t_f(\xi)=0$ and $\ket{\xi}\in PD_f$.\item Noting that $\hs_0=PD_{\epsilon}$, it is obvious.
\item For any $\ket{\psi}\in \hs_0\cap \mathcal{T}_{k}^{-1}(PD_f)$, we have $\tr(\mathcal{T}_{k}(\psi))=\tr(\psi)$ for $\ket{\psi}\in \hs_0$, and $\supp\mathcal{T}_{k}(\psi)\subseteq PD_f$ for $\ket{\psi}\in\mathcal{T}_{k}^{-1}(PD_f)$. Then by Eq.~(\ref{equ:first-n}), we obtain:
    \begin{equation*}\begin{split}0&=t_f(\mathcal{T}_k(\psi))=\tr(\mathcal{T}_k(\psi))-\tr(M_1\mathcal{T}_f(\mathcal{T}_k(\psi))M_1^\dagger)\\
    &=\tr(\psi)-\tr(M_1\mathcal{T}_{kf}(\psi)M_1^\dagger)=t_{kf}(\psi).\end{split}\end{equation*}
    Thus $\ket{\psi}\in PD_{kf}$. It implies that $$\hs_0\cap \mathcal{T}_{k}^{-1}(PD_f)\subseteq PD_{kf}.$$

    Conversely, for any $\ket{\psi}\in PD_{kf}\subseteq \hs_0$, we have $\tr(\mathcal{T}_{k}(\psi))=\tr(\psi)$ and then \begin{equation*}\begin{split}0&=t_{kf}(\psi)=\tr(\psi)-\tr(M_1\mathcal{T}_{kf}(\psi)M_1^\dagger)\\
    &=\tr(\mathcal{T}_{k}(\psi))-\tr(M_1\mathcal{T}_f(\mathcal{T}_k(\psi))M_1^\dagger)=t_f(\mathcal{T}_k(\psi)).\end{split}\end{equation*}
    Therefore, $\ket{\psi}\in \mathcal{T}_{k}^{-1}(PD_f)$. We obtain:
    $$PD_{kf}\subseteq \hs_0\cap \mathcal{T}_{k}^{-1}(PD_f).\ \Box$$\end{enumerate}

We see that $PD_s$ is a subspace of $\hs_0$ for every $s\in S$ by
combining Eq.~(\ref{inter}) and Lemma~\ref{lem:subspace}.

\begin{lemma}\label{lem:inter}
$$PD=\bigcap_{n=0}^\infty PD_n.$$
\end{lemma}

\textit{Proof:} For any state $\ket{\psi}\in PD$, there is some
$s\in S$ such that $$\ket{\psi}\in PD_s\subseteq PD_{s(\leq
n)}\subseteq PD_n$$ for all $n\geq 0$. Thus, $$PD\subseteq \bigcap
PD_n.$$

Conversely, we prove that $\bigcap PD_n\subseteq PD$. Suppose that
$\ket{\psi}\in PD_n$ for all $n\geq 0$. Put $$X=\{f\in S_{fin}|
t_f(\psi)=0\}.$$ Then what we need to do is to find some schedule
$s\in S$ such that $s(\leq n)\in X$ for all $n$. To this end, put
$$E_f=\{g\in X|f\ {\rm is\ a\ prefix\ of}\ g\}$$ for each $f\in
S_{fin}$. We consider the set $$X'=\{f\in S_{fin}| E_f\ {\rm is\ an\
infinite\ set}\}.$$ It holds that $X'\subseteq X$ since $E_f=\emptyset$
for all $f\not\in X$. So, it suffices to find some $s\in S$ such
that $s(\leq n)\in X^\prime$ for all $n$. Now we are going to
construct such a schedule $s$, and our strategy is to define the
head $s(\leq n)$ of $s$ by induction on $n$. First, $s(\leq
0)=\epsilon\in X'$ as $E_\epsilon=X$ is an infinite set. Suppose
that $s(\leq n)=s_1s_2\cdots s_n\in X'$ is already defined. Then
there must be some $s_{n+1}\in\{1,2,\cdots,m\}$ such that $s(\leq
n+1)=s_1s_2\cdots s_ns_{n+1}\in X'$. This is because
$$E_{s(\leq n)}=\{s(\leq n)\}\cup \bigcup_{i=1}^m E_{s(\leq n)i}$$ is an
infinite set, and thus at least one of $E_{s(\leq n)1},E_{s(\leq
n)2},\cdots,E_{s(\leq n)m}$ should be an infinite set. $\Box$

It is also easy to verify that for any $n$, $PD_{n+1}\subseteq
PD_n$. On the other hand, each $PD_n$ is the union of a finite
number of subspaces of $\hs$. The following technical lemma will
help us to further clarify the structure of $PD$.

\begin{lemma}\label{lem:chain}
Suppose that $X_k$ is the union of a finite number of subspaces of
$\hs$ for all $k\geq 1$. If $X_1\supseteq
X_2\supseteq\cdots\supseteq X_k\supseteq\cdots,$ then there exists
$n\geq 1$ such that $X_k=X_n$ for all $k\geq n$.
\end{lemma}

\textit{Proof:} If for some $k\geq 1$, $X_k=\emptyset$, then the result is obvious. So we can assume that $X_k\neq\emptyset$ for all $k\geq 1$. It suffices for us to prove the lemma for a special case that $X_1$ is a single subspace of $\hs$, and then the general case can be obtained by putting $X_0=\hs$ and considering the extended chain $X_0\supseteq
X_1\supseteq\cdots\supseteq X_k\supseteq\cdots.$

Now, we prove the special case by induction on $\dim X_1$. First, for $\dim X_1=0$, $X_k=\{0\}$ for all $k$ and the result holds. For $\dim X_1\geq 1$, we only need to consider the nontrivial case that $X_l\neq X_1$ for some $l$. We choose the minimum one of such $l$, then $X_1=X_2=\cdots=X_{l-1}$ and $X_l$ is a proper subset of $X_1$. Let $X_l=\bigcup_{i=1}^b P_i,$
where $P_1,P_2,\cdots,P_b$ are subspaces of $\hs$. Then for all $k\geq l$,
$$X_k=X_k\cap X_l=\bigcup_{i=i}^b(X_k\cap P_i),$$
and for each $i\in\{1,2,\cdots,b\}$, $P_i$ is a proper subspace of $X_1$ and we have $\dim P_i<\dim X_1$. Therefore, noting that $X_k\cap P_i$ is still a finite union of subspaces, by induction hypothesis, the descending chain
$P_i\supseteq X_{l+1}\cap P_i\supseteq X_{l+2}\cap P_i\supseteq\cdots$
terminates at some $n_i\geq l$, that is $X_k\cap P_i=X_{n_i}\cap P_i$ for all $k\geq n_i$. Let $n=max\{n_i:i=1,2,\cdots,l\}$, then for all $k\geq n$ we have $$X_k=\bigcup_{i=1}^b (X_k\cap P_i)=\bigcup_{i=1}^b (X_n\cap P_i)=X_n.\ \Box$$

Now we can assert that there exists $n\geq 0$ such that $PD_k=PD_n$
for all $k\geq n$, and thus $PD=PD_n$ by combining
Lemmas~\ref{lem:subspace}, \ref{lem:inter} and \ref{lem:chain}.
Indeed, we are able to prove an even stronger result presented in
the following:

\begin{theorem}\label{the:structure} Let $n$ be the smallest integer satisfying $PD_n=PD_{n+1}$. Then $PD=PD_n$.
\end{theorem}

\textit{Proof:} We only need to prove that for any $n\geq 0$,
$PD_n=PD_{n+1}$ implies $PD_{n+1}=PD_{n+2}$. Assume that
$PD_n=PD_{n+1}$ and $\ket{\psi}\in PD_{n+1}$. We are going to show
that $\ket{\psi}\in PD_{n+2}$. By definition, there is
$f=s_1s_2\cdots s_{n+1}\in S_{fin}$ such that $\ket{\psi}\in PD_f$. Put
$f'=s_2s_3\cdots s_{n+1}$. By Eq.~(\ref{equ:recur}), we have
$\supp(\mathcal{T}_{s_1}(\psi))\subseteq PD_{f'}$. On the other
hand, it follows from the assumption that
$$PD_{f'}\subseteq PD_n=PD_{n+1}=\bigcup_{g\in
S_{fin},|g|=n+1}PD_g.$$ Since $PD_{f'}$ and all $PD_g$s are
subspaces of the finite dimensional Hilbert space $\hs$, and a finite-dimensional Hilbert space cannot be the union of its proper subspaces (see Theorem 1.2 of~\cite{Ro07} for reference), there must
be some $g=r_1r_2\cdots r_{n+1}\in S_{fin}$ such that
$PD_{f'}= PD_{f'}\cap PD_g$ and thus
$\supp(\mathcal{T}_{s_1}(\psi))\subseteq PD_g$. We have $\ket{\psi}\in\mathcal{T}_{s_1}^{-1}(PD_g)$. Furthermore, we put
$g'=s_1r_1r_2\cdots r_{n+1}$. Then by Eq.~(\ref{equ:recur}), $\ket{\psi}\in PD_{g'}\subseteq PD_{n+2}$. $\Box$

\section{Quantum Zero-One Law}\label{ZO}

For simplicity of presentation, from now on, we only consider
normalized input state $\rho$, that is, we always assume that
$tr(\rho)=1$.

\begin{definition}The reachable termination probability of program
$\mathcal{P}$ starting in state $\rho$ is the infimum of termination
probability of the program starting in a state reachable from
$\rho$, that is,
$$h(\rho)=\inf\{t(\sigma)|\sigma\in\D(\hs_{R(\rho)}),\tr(\sigma)=1\}.$$\end{definition}

The following lemma gives a characterization of terminating states
in terms of reachable termination probability. It is obviously a
strengthening of Lemma~\ref{lem1}.1.

\begin{lemma}\label{lem:1-condi} $\rho\in T$ (i.e. $t(\rho)=1$) if and only if $h(\rho)=1$.
\end{lemma}

\textit{Proof:} The ``only if" part is obvious. To prove the ``if"
part, we assume that $h(\rho)=1$. Then for any $f\in S_{fin}$, it
follows from Lemma~\ref{lem1}.1 that $t(\mathcal{T}_f(\rho))=1$.
Since $\mathcal{T}_f(\rho)$ can be decomposed as a convex
combination of its eigenvectors, by Lemma~\ref{lem1}.2 we see that
$t(\psi)=1$ whenever $\ket{\psi}$ is an eigenvectors of
$\mathcal{T}_f(\rho).$ We write: $$V_{R(\rho)}=\{{\rm eigenvectors\
of}\ \mathcal{T}_f(\rho)|f\in S_{fin}\}.$$ Then $\hs_{R(\rho)}=\spa
V_{R(\rho)}$, and Lemma~\ref{lem1}.3 implies $t(\psi)=1$ for any
$\ket{\psi}\in\hs_{R(\rho)}$. Finally, for all
$\sigma\in\mathcal{D}(\hs_{R(\rho)})$, since $\sigma$ is a convex
combination of pure states in $\hs_{R(\rho)}$, we assert that
$t(\sigma)=1$ by using Lemma~\ref{lem1}.2 once again. Therefore,
$h(\rho)=1.$ $\Box$

To prove the zero-one law for reachable termination probability, we
need the following technical lemma. It is obvious by definition that
the reachable set is closed under $\mathcal{T}_f$, that is,
$\mathcal{T}_f(R(\rho))\subseteq R(\rho)$ for every $f\in S_{fin}$.
The same conclusion is valid for the reachable space but no so
obvious.

\begin{lemma}\label{fac:inv}
If $\rho\in\D(\hs_{R(\rho)})$, then for any $f\in S_{fin}$,
$\mathcal{T}_f(\rho)\in\D(\hs_{R(\rho)}).$
\end{lemma}

\textit{Proof:} As $\hs_{R(\rho)}$ is finite-dimensional, we can
find a finite subset $F$ of $S_{fin}$ such that
$$\hs_{R(\rho)}=\bigvee_{g\in F}supp(\mathcal{T}_g(\rho)).$$
Thus, for any $\sigma\in\D(\hs_{R(\rho)})$, there exists some
positive real number $\lambda$ such that
$$\sigma\leq\lambda\sum_{g\in F}\mathcal{T}_g(\rho).$$ Let
$$\delta=\lambda\sum_{g\in F}\mathcal{T}_g(\rho)-\sigma\in\D(\hs_{R(\rho)}).$$ Then for any
$k\in\{1,2,\cdots,m\}$,
\begin{equation*}\begin{split}\mathcal{T}_k(\sigma)+\mathcal{T}_k(\delta)&=\mathcal{T}_k(\sigma+\delta)
=\mathcal{T}_k(\lambda\sum_{g\in F}\mathcal{T}_g(\rho))\\
&=\lambda\sum_{g\in
F}\mathcal{T}_{gk}(\rho)\in\D(\hs_{R(\rho)}).\end{split}\end{equation*}
So, we have $\mathcal{T}_k(\sigma)\in\D(\hs_{R(\rho)})$. Moreover,
we obtain $\mathcal{T}_f(\sigma)\in\D(\hs_{R(\rho)})$ for any
$f=k_1\cdots k_m\in S_{fin}$ by induction on $m$. $\Box$

Now we are ready to present the main result in this section.

\begin{theorem}[Zero-One Law]\label{lem:0-1}For any
$\rho$, we have $h(\rho)=0$ or $1$.
\end{theorem}

\textit{Proof:} We write $h=h(\rho)$ and argue that $h>0$ implies
$h=1$. Assume $h>0$. Then for any $\varepsilon>0$, there exists some
$\sigma\in\D(\hs_{R(\rho)})$ such that $\tr(\sigma)=1$ and $h\leq
t_s(\sigma)\leq h+\varepsilon$ for some $s\in S$. We can choose a
sufficiently large integer $n$ such that $t_{s(\leq
n-1)}(\sigma)\geq h/2$ because $\lim_{n\rightarrow\infty}t_{s(\leq
n)}(\sigma)=t_s(\sigma)\geq h.$ Applying Eq.~(\ref{equ:first-n}),
we get: \begin{equation}\label{m1}1-\tr(T_{s(\leq
n)}(\sigma))=t_{s(\leq n-1)}(\sigma)\geq h/2.\end{equation} On the
other hand, we put $\lambda=tr(\mathcal{T}_{s(\leq n)}(\sigma))$.
Then it follows from Lemma~\ref{fac:inv} that
$$\frac{1}{\lambda}\mathcal{T}_{s(\leq
n)}(\sigma)\in\mathcal{D}(\mathcal{H}_{R(\rho)}).$$ Also, it holds
that $tr[\frac{1}{\lambda}\mathcal{T}_{s(\leq n)}(\sigma)]=1$. So,
by the definition of $h(\rho)$ we have
$t(\frac{1}{\lambda}\mathcal{T}_{s(\leq n)}(\sigma))\geq h.$
Consequently,
\begin{equation*}\begin{split}t_{s(>n)}(\mathcal{T}_{s(\leq
n)}(\sigma))&\geq t(\mathcal{T}_{s(\leq n)}(\sigma))=\lambda
t(\frac{1}{\lambda}\mathcal{T}_{s(\leq n)}(\sigma))\\ &\geq\lambda
h= h\cdot\tr(\mathcal{T}_{s(\leq
n)}(\sigma)).\end{split}\end{equation*} Then employing
Eq.~(\ref{equ:two-part}), we obtain:
\begin{equation}\label{m2}\begin{split}h+\varepsilon&\geq t_s(\sigma)=1-\tr(\mathcal{T}_{s(\leq
n)}(\sigma))+t_{s(>n)}(\mathcal{T}_{s(\leq n)}(\sigma))\\ &\geq
1-\tr(\mathcal{T}_{s(\leq n)}(\sigma))+h\cdot\tr(\mathcal{T}_{s(\leq
n)}(\sigma)).\end{split}\end{equation} Now combining Eqs.~(\ref{m1})
and~(\ref{m2}) yields: $$\varepsilon\geq
(1-h)(1-\tr(\mathcal{T}_{s(\leq n)}(\sigma)))\geq
(1-h)\frac{h}{2}.$$ Finally, as $\varepsilon$ can be arbitrarily
small, it holds that $h=1$. $\Box$

From the definition of $h(\rho)$ we see that if $h(\rho)=1$, then
termination probability $t_s(\sigma)=1$ for any state $\sigma$ in
the reachable space $\hs_{R(\rho)}$ of $\rho$ and any schedule $s\in
S$. What happens when $h(\rho)=0$? The following proposition answers
this question.

\begin{lemma}\label{lem:get-state}
If $h(\rho)=0$ then there exists some $\sigma\in\D(\hs_{R(\rho)})$
and $s\in S$ such that $t_s(\sigma)=0$.
\end{lemma}

\textit{Proof:} The proof is divided into three steps. First, we
show that if $h(\rho)=0$ then $t(\sigma)=0$ for some
$\sigma\in\mathcal{D}(\mathcal{H}_{R(\rho)})$. For any two states
$\delta,\theta\in\D(\hs_{R(\rho)})$, and any $s\in S$, we have:
\begin{equation}\label{x}\begin{split}&t(\delta)\leq t_s(\delta)=t_s(\theta)+t_s(\delta-\theta)\\ &\leq
t_s(\theta)+t_s(\sqrt{(\delta-\theta)^2})\leq
t_s(\theta)+\tr\sqrt{(\delta-\theta)^2}\\
& =  t_s(\theta)+\parallel\delta-\theta\parallel_*
 \overset{t_s(\theta)\rightarrow t(\theta)}{\rightarrow}
t(\theta)+\parallel\delta-\theta\parallel_*\end{split}\end{equation}
Since $h(\rho)=0$, we can find a sequence $\{\sigma_n\}$ in
$\D(\hs_{R(\rho)})$ such that $t(\sigma_n)\rightarrow 0\
(n\rightarrow \infty)$. Furthermore, sequence $\{\sigma_n\}$ has an
accumulation point $\sigma$ because $\D(\hs_{R(\rho)})$ is compact
and satisfies the first countability axiom. Thus, there is a
subsequence $\{\sigma_{i_k}\}$ of $\{\sigma_n\}$, which converges to
$\sigma$. It follows from Eq.~(\ref{x}) that $$t(\sigma)\leq
t(\sigma_{i_k})+\parallel\sigma-\sigma_{i_k}\parallel_*\rightarrow
0\ (k\rightarrow\infty).$$

Second, we prove that if $t(\sigma)=0$, then there exists some
$1\leq k\leq m$ such that $t(\mathcal{T}_{k}(\sigma))=0$. It
suffices to see that for any $s\in S$, Eq.~(\ref{equ:two-part})
yields: \begin{equation*}\begin{split}0&=t(\sigma)\geq
t_s(\sigma)=1-\tr(\mathcal{T}_{s_1}(\rho))+t_{s(>1)}(\mathcal{T}_{s_1}(\rho))\\
& \geq t_{s(>1)}(\mathcal{T}_{s_1}(\rho))\geq
t(\mathcal{T}_{s_1}(\rho)) \geq \min_{k=1}^m
t(\mathcal{T}_{k}(\rho))\end{split}\end{equation*}

Third, we show that if $\sigma\in\mathcal{H}_{R(\rho)}$ satisfies
$t(\sigma)=0$ then $t_s(\sigma)=0$ for some schedule $s\in S$. We
recursively construct $s=s_1s_2\cdots\in S$ such that
$t(\mathcal{T}_{s(\leq n)}(\sigma))=0$ for all $n\geq 0.$ For $n=0$,
$t(\mathcal{T}_\epsilon(\sigma))=t(\sigma)=0$. Suppose that
$s_1s_2\cdots s_n$ is already defined and
$t(\mathcal{T}_{s_1s_2\cdots s_n}(\sigma))=0$. Then according to the
conclusion in the above paragraph, we can find $s_{n+1}$ such that
$t(\mathcal{T}_{s_1s_2\cdots
s_ns_{n+1}}(\sigma))=t(\mathcal{T}_{s_{n+1}}(\mathcal{T}_{s_1s_2\cdots
s_n}(\sigma)))=0.$ Finally, we get:
$$t_s(\sigma)=\sum\limits_{n=0}^\infty \tr(M_0 \mathcal{T}_{s(\leq n)}(\sigma)
M_0^\dagger)=0$$ because $\tr(M_0\mathcal{T}_{s(\leq
n)}(\sigma)M_0^\dagger)\leq t(\mathcal{T}_{s(\leq n)}(\sigma))=0$
for all $n\geq 0$. $\Box$

\section{An Algorithm for Termination Checking}\label{Alg}

A combination of the results obtained in Sec.~\ref{RS},~\ref{TS}
and~\ref{ZO} leads to a necessary and sufficient condition for
termination of program $\mathcal{P}$.

\subsection{A Termination Condition}

\begin{theorem}\label{the:main} For any input state $\rho$, $\rho\in
T$ (i.e. $t(\rho)=1)$ if and only if $\hs_{R(\rho)}\cap
PD=\{0\}$.
\end{theorem}

\textit{Proof:} By the zero-one law (Theorem \ref{lem:0-1}) together
with Lemma \ref{lem:1-condi}, we only need to prove that
$h(\rho)=0\ {\rm iff}\ \hs_{R(\rho)}\cap PD\neq\{0\}.$ If
$\hs_{R(\rho)}\cap PD\neq\{0\}$, then we arbitrarily choose
$\ket{\psi}\in \hs_{R(\rho)}\cap PD$,$\ip{\psi}{\psi}=1$ and it holds that
$\psi\in\D(\hs_{R(\rho)})$ and $t(\psi)=0$. Thus, by definition we
have $h(\rho)=0$. Conversely, if $h(\rho)=0$, then it follows from
Lemma \ref{lem:get-state} that there exist
$\sigma\in\D(\hs_{R(\rho)})$ and $s\in S$ with $t_s(\sigma)=0$. Now,
let $\ket{\psi}$ be an eigenvector of $\sigma$. Then $\ket{\psi}\in
\hs_{R(\rho)}$ and $t_s(\psi)=0$. This means that
$\ket{\psi}\in\hs_{R(\rho)}\cap PD\neq\{0\}$. $\Box$

Since we have shown in Sec.~\ref{TS} that $PD$ is a finite union of subspaces, the above condition can be checked by compute the intersections of subspaces pairs for given $\hs_{R(\rho)}$ and $PD$. Therefore, an algorithm for termination checking can be obtained by simply combining an algorithms for computing
reachable states and an algorithm for computing diverging pure
states, which are presented in the next two subsections. An application of these algorithms to checking termination of the example program considered in
Sec.~\ref{sub:qw} is presented in the Appendix.

\subsection{An Algorithm for Computing Reachable States}\label{subs2}
Given a nondeterministic quantum program $\mathcal{P}$ and a initial state $\rho$, Algorithm~\ref{alg:Rs} compute the reachable space $\hs_{R(\rho)}$ based on Theorem~\ref{thm-R}.
\begin{algorithm}
\caption{Computing Reachable States \label{alg:Rs}}
\SetKwInOut{Input}{input}\SetKwInOut{Output}{output}
\Input{An orthonormal basis $B_0$ of $\supp (\rho)$, and a Kraus representation of
$\mathcal{T}(\cdot)=\sum_{j=1}^r E_j\cdot E_j^\dagger$.}
\Output{An orthonormal basis $B$ of $\mathcal{H}_{R(\rho)}$.}
\textbf{set of} states $B\leftarrow\emptyset$\;
(* the number of elements of $B$ *)\\
\textbf{integer} $l\leftarrow 0$\;
(* the index of the state under considering *)\\
\textbf{integer} $i\leftarrow 1$\; %
(* put $B$ to be $B_0$ initially *)\\
\For{$\ket{x}\in B_0$}{
$l\leftarrow l+1$\;
$\ket{b_l}\leftarrow \ket{x}$\;
$B\leftarrow B\cup\{\ket{b_l}\}$\;}
\While{$i\leq l$}{
\For{$j\leftarrow 1$ \KwTo $r$}{
$\ket{x}\leftarrow E_j\ket{b_i}-\sum_{k=1}^l\braket{b_k}{E_j}{b_i}\ket{b_k}$\;
\If{$\ket{x}\neq 0$}{
$l\leftarrow l+1$\;
$\ket{b_l}\leftarrow \ket{x}/\sqrt{\ip{x}{x}}$\;
$B\leftarrow B\cup\{\ket{b_l}\}$\;}}
$i\leftarrow i+1$\;
}
\Return
\end{algorithm}

\textit{Correctness and complexity of the algorithm:} Since $B$ keeps to be a set of
orthonormal states, $l\leq \dim \hs$ always holds during the
execution. Thus, the algorithm terminates after at most $\dim \hs$
iterations of the \textbf{while} loop. Consider any execution of the
algorithm. $B=B_0$ at the beginning, and it is convenient to write
$B_{i-1}$ for the instance of $B$ immediately before the iteration of \textbf{while} loop for $i$. Then $\spa
B_i=\spa B_{i-1}\cup\supp \mathcal{T}(b_i)\subseteq\spa B_{i-1}\cup
\mathcal{T}(\spa B_{i-1}).$ By
Lemma~\ref{reac}, it is easy to prove that $\spa B_i\subseteq
\hs_{\overline{R}_i(\rho)}$ by induction on $i$. Then for the output
$B$, we have $\spa B\subseteq \hs_{R(\rho)}$. On the other hand, we
have $\spa B=\spa B\cup \mathcal{T}(\spa B)$ upon termination of the
algorithm. Then $\hs_{\overline{R}_n(\rho)}\subseteq \spa B$ can be
also proved for all $n$ by induction. Therefore $\hs_{R(\rho)}=\spa
B$.

To get an upper bound of the running time of the algorithm, we write $d=\dim\hs$ and consider each iteration of the \textbf{while} loop: There are $r$ new states $\ket{x}$ being calculated, and each calculation is done in time $O(d^2)$ by multiplying a $d\times d$ matrix and a $d$-dimensional vector. Noting that $r\leq d^2$, the time complexity is $d\cdot r\cdot O(d^2)=O(d^5)$ in total. $\Box$

\subsection{An Algorithm for Computing Diverging Pure States}\label{subs3}
Algorithm~\ref{alg:PD} compute the set of diverging states for a given nondeterministic quantum program. The idea comes from Theorem~\ref{the:structure}:
We calculate $PD_n$ from $PD_{n-1}$, until the condition
$PD_n=PD_{n-1}$ holds. For convenience, we write $J_n=\{PD_f:f\in S_{fin},|f|=n\}$ and thus $\bigcup_{P\in J_n}P=PD_n$. Then to check if $PD_n=PD_{n-1}$, it suffices to check if for any $P\in J_{n-1}$, there exists $Q\in J_n$ such that $P\subseteq Q$.
\begin{algorithm}
\caption{Computing Pure Diverging States \label{alg:PD}}
\SetKwInOut{Input}{input}\SetKwInOut{Output}{output}
\Input{The projection operator of $\hs_0$.}
\Output{A set of subspaces $J_0$.}
(*to record $J_{n-1}$*)\\
\textbf{set of} subspaces $J_0\leftarrow \emptyset$\;
(*to record $J_n$*)\\
\textbf{set of} subspaces $J_1\leftarrow \{\hs_0\}$\;
\textbf{bool} $b\leftarrow 0$\;
\textbf{bool} $c\leftarrow 0$\;
\While{$\neg b$}{
$J_0\leftarrow J_1$\;
$J_1\leftarrow \emptyset$\;
\For{$P\in J_0$}{
\For{$k\leftarrow 1$ \KwTo $m$}{
$J_1\leftarrow J_1\cup \{\mathcal{T}_k^{-1}(P)\cap \hs_0\}$\;}}
(*test if $PD_n=PD_{n-1}$*)\\
$b\leftarrow 1$\;
\For{$P\in J_0$}{
$c\leftarrow 0$\;
\For{$Q\in J_1$}{
$c\leftarrow c\vee(P\subseteq Q)$\;}
$b\leftarrow b\wedge c$\;}}
\Return $J_0$\\
\end{algorithm}

\textit{Correctness of the algorithm:} We prove by induction on $n$ that after the $n$th iteration of \textbf{while} loop, $J_1$ becomes $\{PD_f:f\in S_{fin},|f|=n\}$. For $n=0$, $J_1=\{\hs_0\}=\{PD_\epsilon\}$. Suppose the result holds for $n-1$. At the beginning of the $n$th iteration, $J_0\leftarrow J_1=\{PD_f:f\in S_{fin},|f|=n-1\}$, and then $J_1$ is calculated from $J_0$ by \begin{equation*}\begin{split}J_1&=\{\hs_0\cap\mathcal{T}_k^{-1}(P):1\leq k\leq m,P\in J_0\}\\
&=\{\hs_0\cap\mathcal{T}_k^{-1}(PD_f):1\leq k\leq m,f\in S_{fin},|f|=n-1\}\\
&=\{PD_{kf}:1\leq k\leq m,f\in S_{fin},|f|=n-1\}\\
&=\{PD_f: f\in S_{fin},|f|=n\}.\end{split}\end{equation*}
Here, $PD_{kf}=\hs_0\cap\mathcal{T}_k^{-1}(PD_f)$ comes from Eq.~(\ref{equ:recur}). So we get the correctness of the algorithm. $\Box$

It is worth noting that the termination of this algorithm comes from the descending chain condition in Lemma~\ref{lem:chain}, but the terminating time $n$ is unbounded there. So, it is still unclear how to estimate the number of iterations of the \textbf{while} loop in Algorithm~\ref{alg:PD}, and consequently the complexity of computing the set of diverging states of nondeterministic quantum programs remains unsettled.

\subsection{An example: Quantum Walks}
This subsection is a continuation of Sec.~\ref{sub:qw}. We show how to apply our algorithms developed above to the nondeterministic quantum program $(\{W_1, W_2\},\{P_0,P_1\})$. In Sec.~\ref{sub:qw}, we have shown that this program is not terminating for initial state $\ket{0}$ by observing a diverging path $1212\cdots$. Here, we give an algorithmic check for this fact.

\textit{Computing the reachable space:} We use Algorithm~\ref{alg:Rs} to compute the reachable space $\hs_{R(\op{0}{0})}$. The Kraus operators of $\mathcal{T}(\cdot)$ are $E_1=W_1P_1$ and $E_2=W_2P_1$. We write $B_{i,j}$ and $\ket{x_{i,j}}$ for the instance of $B$ and $\ket{x}$ respectively, for index $i$ and index $j$ during the execution. Then $B$ is calculated by a finite number of iterations of \textbf{while} loop as follows:\\
\indent Initially we have
\begin{equation*}\begin{split}
 l&=1,\\
\ket{b_1}&=\ket{0},\\
B_0&=\{\ket{b_0}\};
\end{split}\end{equation*}
\indent for the iteration of $i=1$,\\
\indent\indent for $j=1$,
\begin{equation*}\begin{split}
E_1\ket{b_1}&=(\ket{0}+\ket{1}+\ket{3})/\sqrt{3},\\
\ket{x_{1,1}}&=(\ket{1}+\ket{3})/\sqrt{3}\neq 0,\\
l&=1+1=2,\\
\ket{b_2}&=(\ket{1}+\ket{3})/\sqrt{2},\\
B_{1,1}&=\{\ket{b_1},\ket{b_2}\};
\end{split}\end{equation*}
\indent\indent for $j=2$,
\begin{equation*}\begin{split}
E_2\ket{b_1}&=(\ket{0}-\ket{1}+\ket{3})/\sqrt{3},\\
\ket{x_{1,2}}&=(-\ket{1}+\ket{3})/\sqrt{3}\neq 0,\\
l&=2+1=3,\\
\ket{b_3}&=(-\ket{1}+\ket{3})/\sqrt{2},\\
B_{1,2}&=\{\ket{b_1},\ket{b_2},\ket{b_3}\};
\end{split}\end{equation*}
\indent for the iteration of $i=2$,\\
\indent\indent for $j=1$,
\begin{equation*}\begin{split}
E_1\ket{b_2}&=(-\ket{1}+2\ket{2}+\ket{3})/\sqrt{6},\\
\ket{x_{2,1}}&=2\ket{2}/\sqrt{6}\neq 0,\\
l&=3+1=4,\\
\ket{b_4}&=\ket{2},\\
B_{2,1}&=\{\ket{b_1},\ket{b_2},\ket{b_3},\ket{b_4}\};
\end{split}\end{equation*}
\indent\indent for $j=2$,
\begin{equation*}\begin{split}
E_2\ket{b_2}&=(2\ket{0}+\ket{1}-\ket{3})/\sqrt{6},\\
\ket{x_{2,2}}&=0;
\end{split}\end{equation*}
\indent for the iteration of $i=3$,\\
\indent\indent for $j=1$,
\begin{equation*}\begin{split}
E_1\ket{b_3}&=(-2\ket{0}+\ket{1}+\ket{3})/\sqrt{6},\\
\ket{x_{3,1}}&=0;
\end{split}\end{equation*}
\indent\indent for $j=2$,
\begin{equation*}\begin{split}
E_2\ket{b_3}&=(-\ket{1}-2\ket{2}-\ket{3})/\sqrt{6},\\
\ket{x_{3,2}}&=0;
\end{split}\end{equation*}
\indent for the iteration of $i=4$,\\
\indent\indent for $j=1$,
\begin{equation*}\begin{split}
E_1\ket{b_4}&=0,\\
\ket{x_{4,1}}&=0;
\end{split}\end{equation*}
\indent\indent for $j=2$,
\begin{equation*}\begin{split}
E_2\ket{b_4}&=0,\\
\ket{x_{4,2}}&=0.
\end{split}\end{equation*}

So the output is
$$B=\{\ket{0},(\ket{1}+\ket{3})/\sqrt{2},(-\ket{1}+\ket{3})/\sqrt{2},\ket{2}\},$$
and the reachable space $\hs_{R(\op{0}{0})}=\spa B$ is actually the whole state space.

\textit{Computing the set of pure diverging states:} We use Algorithm~\ref{alg:PD} to compute the set of pure diverging states $PD$. In the algorithm, $PD_f$ is recursively calculated by Eq.~(\ref{equ:recur}). Specifically, here we have $$\T_k(\cdot)=W_kP_1\cdot(W_kP_1)^\dagger\ (k=1,2)$$
and then the projection operator of
$\hs_0\cap\T_k^{-1}(P)$ is exactly $P_1\cap W_k^{-1}P W_k$. Now, we calculate each $PD_f$ recursively on $|f|$ as follows:\\
\indent For $|f|=0$, we initially have\\
$$PD_\epsilon=P_1=Id_4-\op{2}{2};$$
\indent for $|f|=1$,\\
\indent\indent to compute $PD_1$ we get that
\begin{equation*}\begin{split}
&W_1^{-1}\ket{2}=(\ket{1}+\ket{2}+\ket{3})/\sqrt{3},\\
&W_1^{-1}PD_\epsilon W_1=\{W_1^{-1}\ket{2}\}^\perp,
\end{split}\end{equation*}
\indent\indent then\\
$$PD_1=P_1\cap W_1^{-1}PD_\epsilon W_1=\op{0}{0}+\op{-}{-},$$
\indent\indent where $\ket{-}=(\ket{1}-\ket{3})/\sqrt{2}$;\\
\indent\indent to compute $PD_1$ we get that\\
\begin{equation*}\begin{split}
&W_2^{-1}\ket{2}=(\ket{1}+\ket{2}-\ket{3})/\sqrt{3},\\
&W_2^{-1}PD_\epsilon W_2=\{W_2^{-1}\ket{2}\}^\perp,
\end{split}\end{equation*}
\indent\indent then\\
$$PD_2=P_1\cap W_2^{-1}PD_\epsilon W_2=\op{0}{0}+\op{+}{+},$$
\indent\indent where $\ket{+}=(\ket{1}+\ket{3})/\sqrt{2}$;\\
\indent for $|f|=2$,\\
\indent\indent to compute $PD_{11}$ we get that\\
\begin{equation*}\begin{split}
&W_1^{-1}\ket{0}=(\ket{0}+\ket{1}-\ket{3})/\sqrt{3},\\
&W_1^{-1}\ket{-}=(-\ket{1}+2\ket{2}-\ket{3})/\sqrt{6},
\end{split}\end{equation*}
\indent\indent then\\
$$PD_{11}=(\ket{0}+\ket{1}-\ket{3})(\bra{0}+\bra{1}-\bra{3})/3;$$
\indent\indent to compute $PD_{21}$ we get that\\
\begin{equation*}\begin{split}
&W_2^{-1}\ket{0}=(\ket{0}+\ket{1}+\ket{3})/\sqrt{3},\\
&W_2^{-1}\ket{-}=(-2\ket{0}+\ket{1}+\ket{3})/\sqrt{6},
\end{split}\end{equation*}
\indent\indent then\\
$$PD_{21}=P_1\cap W_2^{-1}PD_1 W_2=\op{0}{0}+\op{+}{+};$$
\indent\indent to compute $PD_{12}$ we get that\\
\begin{equation*}\begin{split}
&W_1^{-1}\ket{0}=(\ket{0}+\ket{1}-\ket{3})/\sqrt{3},\\
&W_1^{-1}\ket{+}=(2\ket{0}-\ket{1}+\ket{3})/\sqrt{6},
\end{split}\end{equation*}
\indent\indent then\\
$$PD_{12}=P_1\cap W_1^{-1}PD_1 W_1=\op{0}{0}+\op{-}{-};$$
\indent\indent to compute $PD_{22}$ we get that\\
\begin{equation*}\begin{split}
&W_2^{-1}\ket{0}=(\ket{0}+\ket{1}+\ket{3})/\sqrt{3},\\
&W_2^{-1}\ket{+}=(\ket{1}-2\ket{2}-\ket{3})/\sqrt{6},
\end{split}\end{equation*}
\indent\indent then\\
$$PD_{22}=(\ket{0}+\ket{1}+\ket{3})(\bra{0}+\bra{1}+\bra{3})/3.$$

Since $PD_1=PD_{12}$ and $PD_2=PD_{21}$, we have $PD=PD_1\cup PD_2$.

Finally, we get that
$$\hs_{R(\op{0}{0})}\cap PD=\spa\{\ket{0},\ket{-}\}\cup\spa\{\ket{0},\ket{+}\}\neq \{0\}.$$
So, this program is not terminating.

\section{Conclusion}\label{Con}
In this paper, we defined a mathematic model of nondeterministic
quantum programs, in which a program consists of a collection of
quantum processes, each process is represented by a quantum Markov
chain over the common state space, and the execution of these
processes are nondeterministically scheduled. The advantage of this
model is that it is independent of the details of its
implementations so that we can focus our attention on examining
high-level behaviors of nondeterministic quantum programs. In
particular, a termination condition for nondeterministic quantum
programs was found, and a classical (not quantum) algorithm for
their termination checking was designed. To achieve these results,
several new mathematical tools have been developed to attack the
difficulty arising from the combined complexity of quantum setting
and nondeterminism:
\begin{itemize}
\item We established a quantum zero-one law for termination
probability of nondeterministic quantum programs. This law allows us
to reduce the termination checking problem to emptiness checking of
the intersection of the reachable space and the space of diverging
pure states, instead of calculating the terminating probabilities
over infinitely many execution schedules.
\item We found an equivalence between the reachable space of a
collection of super-operators and that of their arithmetic average.
\item It was shown that the descending chain condition holds for finite unions of subspaces of a finite-dimensional
Hilbert space. This helps us to extend our proof techniques for a
single subspace to the case of multiple subspaces, which are
unavoidable when nondeterministic choices are present.
\end{itemize}

For the further studies, an immediate topic is to extend the results
presented in this paper to quantum concurrent programs where not all
but only fair execution schedules are allowed. A major difficulty
for such an extension comes from an essential difference between
quantum concurrent programs and classical (and probabilistic)
concurrent programs. In the classical case, the behavior of a
concurrent program can be visualized as a directed transition graph,
in which only an ordering structure determined by transition
relation exists. In the state space of a quantum concurrent program,
however, a linear algebraic structure and a transition relation
lives together. Those methods of searching in the state space of a
classical (and probabilistic) concurrent program developed in the
literature (see for example~\cite{HSP83}) are not effective in the
quantum case because they usually violate the linear algebraic
structure of the state space of a quantum program. It seems that a
new theory of quantum graphs, where their linear algebraic and
ordering structures are coordinated well, is essential for the
studies of quantum concurrent programs.

\section*{Acknowledgment}
We are grateful to Runyao Duan and Yuan Feng for useful discussions.
This work was partly supported by the Australian Research Council
(Grant No: DP110103473) and the National Natural Science Foundation
of China (Grant No: 60736011).


\begin{thebibliography}{99}

\bibitem{Ab04} S. Abramsky, High-level methods for quantum computation
and information, in: \textit{Proceedings of the 19th Annual IEEE
Symposium on Logic in Computer Science (LICS)}, 2004, pp. 410-414.

\bibitem{AAKV01} D. Aharonov, A. Ambainis, J. Kempe and U. V.
Vazirani, Quantum walks on graphs, in: \textit{Proceedings on 33rd
Annual ACM Symposium on Theory of Computing (STOC)}, 2001, pp.
50-59.

\bibitem{AG05} T. Altenkirch and J. Grattage, A functional quantum
programming language, in: \textit{Proceedings of the 20th Annual
IEEE Symposium on Logic in Computer Science (LICS)}, 2005, pp.
249-258.

\bibitem{BS06} A. Baltag and S. Smets, LQP: the dynamic logic of quantum
information, \textit{Mathematical Structures in Computer Science},
16(2006)491-525.

\bibitem{BCS03} S. Betteli, T. Calarco and L. Serafini,
Toward an architecture for quantum programming, \textit{European
Physics Journal}, D25(2003)181-200.

\bibitem{BJ04} O. Brunet and P. Jorrand, Dynamic quantum logic for
quantum programs, \textit{International Journal of Quantum
Information}, 2(2004)45-54.

\bibitem{CMS06} R. Chadha, P. Mateus and A. Sernadas, Reasoning about imperative quantum programs,
\textit{Electronic Notes in Theoretical Computer Science},
158(2006)19-39.

\bibitem{DP06} E. D'Hondt and P. Panangaden, Quantum weakest preconditions,
\textit{Mathematical Structures in Computer Science},
16(2006)429-451.

\bibitem{FDJY07} Y. Feng, R. Y. Duan, Z. F. Ji and M. S. Ying, M, Proof rules
for purely quantum programs, \textit{Theoretical Computer Science}
386(2007) 151-166.

\bibitem{FDY11} Y. Feng, R. Y. Duan and M. S. Ying, Bisimulation for quantum
processes, in: \textit{Proceedings of the 38th ACM SIGPLAN-SIGACT
Symposium on Principles of Programming Languages (POPL)}, 2011, pp.
523-534.

\bibitem{G06} S. J. Gay, Quantum programming languages: survey and bibliography,
\textit{Mathematical Structures in Computer Science},
16(2006)581-600.

\bibitem{GN05} S. J. Gay and R. Nagarajan, Communicating quantum
processes, in: \textit{Proceedings of the 32nd ACM SIGPLAN-SIGACT
Symposium on Principles of Programming Languages (POPL)}, 2005, pp.
145-157.

\bibitem{HSP83} S. Hart, M. Sharir and A. Pnueli, Termination of
probabilistic concurrent programs, \textit{ACM Transactions on
Programming Languages and Systems}, 5(1983)356-380.

\bibitem{JL04} P. Jorrand and M. Lalire, Toward a quantum process
algebra, in: \textit{Proceedings of the First ACM Conference on
Computing Frontiers}, 2004. pp. 111-119.

\bibitem{MJ01} G. Mitchison and R. Josza, Counterfactual
computation, \textit{Proceedings of the Royal Society of London A},
457(2001)1175-1193.

\bibitem{NPW07} R. Nagarajan, N. Papanikolaou and D. Williams,
Simulating and compiling code for the sequential quantum random
access machine, \textit{Electronic Notes in Theoretical Computer
Science}, 170(2007)101-124.

\bibitem{Om03} B. \"{O}mer, \textit{Structured quantum programming},
Ph.D thesis, Technical University of Vienna (2003)

\bibitem{SZ00} J. W. Sanders and P. Zuliani, Quantum programming,
in: \textit{Proceedings, Mathematics of Program Construction 2000},
LNCS 1837, pp. 80-99.

\bibitem{Se04} P. Selinger, Towards a quantum programming language,
\textit{Mathematical Structure in Computer Science},
14(2004)527-586.

\bibitem{Se04a} P. Selinger, A brief survey of quantum programming
languages, in: \textit{Proceedings of the 7th International
SymposiumFunctional and Logic Programming (FLOPS)}, 2004, pp. 1-6.

\bibitem{SA06} K. M. Svore, A. V. Aho, A. W. Cross, I. Chuang and
I. L. Markov, A layered software architecture for quantum computing
design tools, \textit{IEEE Computer}, 39(2006)58-67.

\bibitem{Yi11} M. S. Ying, Floyd-Hoare logic for quantum programs,
\textit{ACM Transactions on Programming Languages and Systems}
(accepted).

\bibitem{YDFJ10} M. S. Ying, R. Y. Duan, Y. Feng and Z. F. Ji, Predicate transformer semantics of quantum programs.
in: S. Gay and I. Mackie (eds.), \textit{Semantic Techniques in
Quantum Computation}, Cambridge University Press, 2010, pp. 311-360.

\bibitem{YF10} M. S. Ying and Y. Feng, Quantum loop programs, \textit{Acta Informatica} 47(2010)221-250.

\bibitem{YF11} M. S. Ying and Y. Feng, A flowchart language for quantum programming,
\textit{IEEE Transactions on Software Engineering}, 37(2011)466-485.

\bibitem{Zu04} P. Zuliani, Nondeterminstic quantum programming, In:
P. Selinger (ed.), \textit{Proceddings of the 2nd International
Workshop on Quantum Programming Languages}. pp. 179-195.

\bibitem{Zu05} P. Zuliani, Compiling quantum programs, \textit{Acta
Informatica}, 41(2005)435-473.

\bibitem{Ro07} S. Roman, \textit{Advanced Linear Algebra}, Springer Press, 2008.
\end{thebibliography}
\end{document}